\documentclass[12pt,a4paper]{article}
\usepackage{amsmath,amssymb,epsfig}

\newcommand{\be}{\begin{equation}\label}
\newcommand{\ee}{\end{equation}}
\newcommand{\prt}{\partial}

\begin{document}

\textwidth=135mm
 \textheight=200mm
\begin{center}
{\bf \Large Gauge origin of the Dirac field \\
and singular solutions to the Dirac equation
}
\vskip 5mm
Vladimir V. Kassandrov 
\vskip 5mm
{\it Institute of gravitation and cosmology, \\ Peoples' Friendship University of Russia}  
\end{center}

\section {On  the well and not so well known: relativistic field equations and their solutions}

The close relationship established in the framework of relativistic field theory between the physical space-time Minkowski geometry $\bf M$ and the two observed types of elementary particles, {\it bosons} and {\it fermions}, is, certainly, one of the most fundamental. The connection manifests itself in the existence, discovered by E. Cartan~\cite{Cartan}, of two and only two types of irreducible representations of the Lorentz group, symmetry group of $\bf M$, -- {\it tensorial} and {\it spinorial} ones. According to the modern paradigm, each type of particles of {\it integer} and {\it half-integer} spins (bosons and fermions, respectively) is described by some linear field equation form-invariant under transformations of the Lorentz group. To ensure the form-invariance, the fields themselves (``wave functions'') should transform through irreducible representations of the Lorentz group (bosons through tensorial while fermions through spinorial ones, respectively). 

Undoubtedly, this rule is fulfilled for particles of low spins. Specifically, spinless bosons ($\pi$-mesons) are described by scalar fields subject to the Klein-Gordon equation whereas unit spin particles -- to the Proca equations for massive vector field or wave equations for 4-potentials of massless electromagnetic field. Fundamental fermions (electron, neutrino, proton, neutron, etc.) of spin $1/2$ are described by the Dirac equation whose 4-component field transforms as a bispinor (a pair of 2-spinors)~\footnote{Massless fermions should be described by the Weyl equation. However, after the concept of {\it massive} neutrino has gained wide acceptance, the corresponding entry in the table of fundamental particles turned out to be vacant, making the situation look rather strange from a general viewpoint}. 

The Dirac equation corresponds to the principal constituents of matter. Introduction of interaction between fermions as matter fields is then realized via the fields of integer spin, carriers of interaction, on the base of the requirement of local gauge invariance of the full Lagrangian. Under gauge transformations the local phase of a wave function and the potentials of a gauge field change concordantly (the latter in a gradient-wise way) while the field strengths remain invariant. However, not all of the integer spin fields can be considered as gauge fields; this, in particular, is true for the Klein-Gordon ``mesonic'' field.  In the same manner, the Dirac field is not a gauge one by itself. {\it This  seemingly evident statement will, however, be disputed further on}. 

Let us now consider the  {\it solutions} of the equations for free relativistic fields. Apart from the most often encountered {\it everywhere regular} solutions of the plane wave type, there exists a wide class of these solutions {\it singular} on a zero measure set, with isolated pointlike or string-like singularities. Besides spherical waves, well known examples of solutions with a pointlike singularity are the Coulomb solution to Maxwell equations or the mesonic Yukawa potential subject to the Klein-Gordon equation. Not always one can put in correspondence with such a singular solution some $\delta$-like {\it source}: for example, this is impossible for the ``flat limit'' electromagnetic field of the Kerr-Newman solution in GTR (the so-called Appel solution~\cite{Witt}) with a ring-like singularity, because of its twofold structure. 

It is also known that in relativistic QM wave functions of the bound $s$- and $p$- states of the hydrogen atom have a weak singularity in the origin~\cite{Davydov,Fock}. It turns out that solutions with not only point- or string- but even membrane-type singularities do exist for free Maxwell, Yang-Mills and Weyl fields~\cite{IJGMMP}. As we shall show later, the massive Klein-Gordon and Dirac fields  are no exception. It is this, most general class of solutions to relativistic field equations that we shall deal with below. Some of such singular solutions are well known and possess generally accepted physical interpretation; others seem to be new. 

In the case when {\it all the solutions} to a fundamental field equation can be obtained from solutions of another equation, and {\it vice versa}, one should not consider such equations as independent and describing {\it different} types of particles. On the contrary, these equations should, perhaps, be treated as mathematically equivalent and corresponding to one and the same physical system in different representations. An evident example of such a situation is  the link between Maxwell equations for field strengths and wave equations for electromagnetic potentials. Indeed, one can (at least locally) assign to any solution of Maxwell equations  a class of (gauge equivalent) potentials subject to wave equations, and vice versa. That is why we, of course, do not associate these equations with different physical entities but consider them both as describing one and the same electromagnetic field in different representations. 

It turns out that a similar {\it equivalence} relation exists between the solutions to massless ( Maxwell, Weyl and d'Alembert)~\cite{Sing,Equiv} and massive (Dirac and Klein-Gordon)~\cite{Equiv,Zommer} equations describing {\it ostensibly different} types of particles. Evidently, the present situation is at discord with the generally accepted viewpoint. Indeed, in the case of, say, massive fields it is usually postulated that the Dirac equation (DE) and the equation of Klein-Gordon (KGE) are responsible for description of particles of different spins, possess different sets of conserved quantities and transform  through different representations of the Lorentz group. As to  their solutions, everybody knows that each component of the Dirac field identically satisfies the KGE, but the converse is false! In this sense the DE is usually considered more {\it rigid} and informative than the KGE ~\cite{Bogol}. 

Meanwhile, the correspondence between the solutions to the DE and KGE can be non-algebraic but {\it differential} in nature just it is the case between Maxwell equations  and the wave equations for electromagnetic potentials. Indeed, as it was proved in ~\cite{Equiv} (see also~\cite{Zommer}), {\it any solution to the DE can be obtained by differentiation of a corresponding set of four solutions to the KGE} (defined up to a specific ``gauge'' freedom, see below).

In other words, one can regard a quadruple of the Klein-Gordon fields as a sort of ``potentials'' for the Dirac bispinor field~\cite{Equiv}. From this point of view the {\it Dirac field is itself a gauge field},  and the DE and KGE should be considered as mathematically equivalent and describing one and the same type of particles. The  corresponding construction proposed earlier in ~\cite{Equiv} is presented in section 2. In section 3, by making use of the 2+2 representation of the DE,  this construction is refined and essentially reinforced. It is shown, in particular, that only a pair (instead of  the entire quadruple) of the solutions to the KGE is sufficient to obtain through differentiation an arbitrary solution to the DE.

Since the components of the arising solutions to the DE, in turn, also satisfy the KGE, it becomes possible to generate whole {\it chains} of the DE-KGE solutions. Such a possibility is demonstrated in section 4 on a number of examples for which, as the starting points, two static spherically symmetric solutions to the KGE are taken (one of them being the mesonic Yukawa potential) as well as a number of stationary axisymmetric solutions. The chain of the DE-KGE solutions arising in the first case resembles multipole harmonics for the massless fields and could, perhaps, have nontrivial physical interpretation. Particularly, a ``spinorial analogue'' of the Yukawa potential is obtained. Solutions of the second chain possess a singularity of the ``Dirac string'' type.   

On the other hand, the discovered possibility to obtain {\it general solution to the DE} from {\it scalar} fields poses, in the framework of the considered approach, the question about the origin of the {\it spinor} law of transformation of the Dirac field.  A  possible solution of this problem was proposed in ~\cite{Equiv}.  It is based on the use of {\it internal symmetry} of the KGE system with respect to transformations of the group $SL(4,\mathbb C)$ {\it intermixing} components of the quadruple of the Klein-Gordon ``potentials''. After appropriate ``tuning'' of such transformations to the transformations of the Lorentz group the canonical spinor law of transformations of the Dirac field can be completely restored.  Generally, however, the arising law of transformations of the Dirac field components corresponds to a {\it nonlinear} representation of the Lorentz group and, remarkably, does not result in a spinorial two-valuedness. In particular, {\it after a full rotation the Dirac field returns, as a rule,  back to its original value}. The properties of admissible transformations of the DE-KGE solutions are examined in section 5 and examplified on the solutions from section 4.  

Finally, in section 6 we examine the problem of ``twofoldness'' of the set of conserved quantities which can be associated with any solution of both the DE and the KGE owing to their mutual correspondence. Primarily, it can be proved that there exist two different ``energies'' of the Dirac field one of which, according to the properties of the corresponding scalar fields, is positive definite! Conversely, to any solution of the KGE a positive definite ``probability density'' can be ascribed! There are also two admissible expressions for angular momenta of the DE-KGE solutions.

In section 7 we briefly describe the situation arising in attempting to generalize the elaborated construction to the case when an external electromagnetic field is present. In conclusion (section 8) the most important results of the paper are summarized. The problem of appropriate physical interpretation of the established equivalence of distinct relativistic field equations is touched upon as well as the consequences of this  equivalence for QFT.  

To keep things simple, in the main part of the paper we do not apply the 2-spinor formalism but use instead an equivalent 2+2 matrix form of representation. For the metric on $\bf M$ the form  $\eta_{\mu\nu}=diag\{+1,-1,-1,-1\}$ is chosen so that, say, the d'Alembert operator has the form  $\Box:=-\partial_\mu\partial^\mu=\Delta - \partial^2/\partial t ^2$. As it is the custom, the system of units where $c=1,\hbar=1$ is used throughout the paper.

\section{The Klein-Gordon ``potentials'' and gauge invariance of the free Dirac equation}

Consider the Klein-Gordon equation (KGE)
\be{column}
(\Box - m^2)\phi = 0 ,
\ee
for four free complex scalar fields  $\phi=\{\phi_a\},~~a=1,2,3,4$. 
The Klein-Gordon operator can be factorized 
\be{factor}
(\Box - m^2) = DD^* = D^*D 
\ee
into the product of two commuting Dirac operators $D,D^*$ of the first order: 
\be{oper}
D:= i\gamma^\mu \prt_\mu - m,~~~D^*:=i\gamma^\mu \prt_\mu + m,
\ee
where $\gamma^\mu$  are $4\times 4$  Dirac matrices 
\be{commut}
\gamma^\mu \gamma^\nu + \gamma^\nu \gamma^\mu = 2\eta^{\mu\nu}, 
\ee
which we shall take in the standard 2+2 representation making use of the Pauli matrices  (see below, section 3).

Through the derivatives of $\phi$ let us then define another 4-component complex field  $\chi$,  
\be{poten}
\chi:=D^*\phi, 
\ee
which, by (\ref{column}) and (\ref{factor}), defines a solution to the Dirac equation (DE),  
\be{dirac}
D\chi = DD^*\phi = 0.
\ee

Conversely, let an arbitrary solution of the DE $\chi$ be given (it is then known that each component of $\chi$ identically satisfies the KGE,  since $0=D^*(D\chi) = (\Box-m^2)\chi = 0$). In this case, the system of four inhomogeneous first order equations (\ref{poten}) can be always (locally) resolved with respect to four unknowns $\phi$ (this will be explicitly demonstrated below, see section 3). Of course, the obtained functions $\phi$ are subject to the KGE,
\be{ident2}
0=D\chi = DD^*\phi=(\Box-m^2)\phi \equiv 0, 
\ee
yet determined not uniquely but up to a general solution of an {\it homogeneous} equation of the type (\ref{poten}). Specifically, any given solution $\chi$ to the DE  (``strengths'' of the Dirac field) stays invariant under the following {\it gauge} transformations of the corresponding ``Klein-Gordon potentials'' from (\ref{poten}): 
\be{gauge}  
\phi \mapsto \phi + \kappa, 
\ee
with $\kappa$ being some {\it arbitrary} solution of the (conjugate) DE, 
\be{dirac2}
D^*\kappa = 0. 
\ee
Now, since for each $\kappa$ some Klein-Gordon potentials $\xi$ exist, that is $\kappa = D\xi$, the gauge transformation  ({\ref{gauge}) can be represented in a familiar gradient form
\be{grad}
\phi \mapsto \phi + D\xi. 
\ee
Thus, for any DE solution potentials subject to the KGE are locally defined up to the gauge transformations (\ref{grad}), and through their differentiations {\it a whole set of solutions to the free DE can be obtained}~\cite{Equiv}. As to the DE itself, {\it it should be regarded as a gauge field} analogous to Maxwell equations for the strengths of electromagnetic field.

However, the gauge function $\xi$, in contrast with the gauge symmetry in electrodynamics, is not arbitrary but subject to the KGE, $(\Box-m^2)\xi =0$. This resembles the ``residual'' gauge invariance
\be{maxgauge}
A_\mu \mapsto A_\mu -\prt_\mu \alpha,~~~~\Box \alpha =0
\ee
of Maxwell equations $\prt^\nu F_{\mu\nu}=0,~F_{\mu\nu}=\prt_\mu A_\nu-\prt_\nu A_\mu$, {\it supplemented by the Lorentz equation for the potentials  $\prt_\mu A^\mu =0$}. 

It is  worth noting that similar ``weak'' gauge invariance (for which the gauge function can depend on coordinates {\it implicitly}, only through the components of the field function under transform) holds for the class of solutions to relativistic field equations generated by twistor functions (for detail, see~\cite{IJGMMP}).

\section{Any solution to the Dirac equation from a doublet of the Klein-Gordon scalar fields}

We now intend to enhance the above elucidated results of the paper~\cite{Equiv} and  demonstrate that at most {\it two} solutions to the KGE are sufficient to obtain, by differentiation, any solution to the DE
\be{Dirac2}
D\psi: = (i\gamma^\mu \prt_\mu -m)\psi =0.
\ee
To this end, let us consider the well-known matrix 2+2 ``split'' form of the DE (see, e.g.,~\cite[ch.II, sect.9]{Akhiezer}):
\be{Dirac2+2}
W a = -\imath m b, ~~ \tilde W b = - \imath m a, 
\ee
with matrix-valued operators
\be{weyl}
W:=(\prt_t - \vec \sigma \nabla), ~~ \tilde W:=(\prt_t + \vec \sigma \nabla)
\ee
which we shall call {\it Weyl operators} (principal and conjugate, respectively). Here $\vec \sigma =\{\sigma_a\},~a=1,2,3$  are the Pauli matrices, the 2-spinors $\{a,b\}$ are defined as the half-sum/half-difference of the initial Dirac 2-spinors $\psi^T =\{\kappa,\chi\}$, 
\be{2-spinors}
a=(\kappa+\chi)/2,~~~b=(\kappa-\chi)/2 ,  
\ee 
and Weyl operators $W,\tilde W$ factorize the d'Alembert wave operator 
\be{dalamber}
W\tilde W = \tilde W W=-\Box =\prt^2/\prt t^2 - \Delta. 
\ee 
The above described procedure of seeking the Klein-Gordon potentials based on resolving  equations (\ref{poten}), in the 2+2 representation reduces to the following system of equations 
\be{poten2+2}
a = \tilde W \beta - \imath m \alpha, ~~~ b = W \alpha - \imath m \beta
\ee
with respect to a pair of unknown 2-component functions $\{\alpha,\beta\}$ for any given 2-spinors $\{a,b\}$ subject to the DE (\ref{Dirac2+2}). It is easy to check that if the solution of (\ref{poten2+2}) exists, it is non-unique, and that the potentials $\{\alpha,\beta\}$, on account of (\ref{Dirac2+2}) and (\ref{dalamber}), should satisfy the KGE, $(\Box-m^2)\alpha=0$,~~$(\Box-m^2)\beta =0$. Specifically, the gauge transformation of potentials which leave invariant both Dirac 2-spinors has the form 
\be{gauge2+2}
\begin{array}l
\alpha \mapsto \alpha -m^2 \pi - \imath m \tilde W \rho, \\
\beta \mapsto \beta - m^2 \rho - \imath m W \pi. 
\end{array}
\ee
Here $\pi$ and $\rho$ are two arbitrary and independent pairs of functions each component of which satisfies the KGE.

Now the problem of {\it existence} of potentials subject to  (\ref{poten2+2}) can be explicitly resolved. Indeed, let us nullify one of the 2-component potentials setting, say, $\beta=0$.   Then the system (\ref{poten2+2}) reduces to {\it identification} of the other 2-component potential function $\alpha$ with the first of the given 2-spinors 
\be{first2sp}
\alpha = \frac{\imath}{m} a, 
\ee
while the other 2-spinor is then expressed through the derivatives of the first one:
\be{second2sp}
b=\frac{\imath}{m} W a. 
\ee   

In other words, for any solution of the DE the first of equations (\ref{poten2+2}) is simply a definition ({\ref{second2sp}) of the second 2-spinor $(b)$ through the first one $(a)$, after which the second equation is identically satisfied since the KGE holds for both components of the principal 2-spinor $(a)$. 

Thus, {\it any solution to the DE is given by some two functions $(a)$ subject to the KGE} and defining the first Dirac 2-spinor whereas the second 2-spinor $(b)$ is explicitly expressed through the derivatives of the first one $(a)$! In the next section we shall exhibit simple examples of such a procedure and obtain a number of singular solutions to the DE.  
 
\section{Chains of singular solutions to the Dirac and Klein-Gordon equations}
 
In the above described method of generation of the DE solutions from some pair of the KGE solutions, components of the  obtained Dirac fields, as usual, satisfy the KGE themselves and, therefore, can serve as ``potentials'' to obtain new solutions to the DE, and so on. The chain of the DE-KGE solutions thus arising proves to be infinite or terminates in the case when new solutions happen to be functionally dependent on the previous ones. 
 
Because of the {\it linearity} of the considered equations it is sufficient to restrict oneself to the case when only one of the generating KGE solutions is nonzero and take, say, the initial 2-spinor in the form $a^T = (0, F)$ or $a^T=(G,0)$, where the functions $F,G$  represent some solutions to the KGE. The general case is then given by the 2-spinor $a^T=(G,F)$, which is a superposition of the DE solutions generated by ``partial'' 2-spinors.  
 
Consider now a simple yet important example of a chain of singular solutions to the DE starting from the ``mesonic''  Yukawa potential of the form $-g^2 \exp(-mr)/r$, which is a static and spherically symmetric solution to the KGE. Namely, let us take (removing for simplicity the scale factor)
\be{yukava_a}
a=\left(\begin{array}l

0 \\ 
F
  \end{array}
  \right), 
  ~~~ F=\frac{1}{r}e^{-mr}.
 \ee
Making now use of  (\ref{second2sp}) and computing  the conjugate 2-spinor, we get (in the Cartesian coordinates $\{x,y,z\},~r:=\sqrt{x^2+y^2+z^2}$): 
\be{yukava_b}
 b=\frac{\imath}{mr^3}(1+mr)e^{-mr}\left(\begin{array}l
~~w\\ 
-z
  \end{array}
  \right),~~~ w:=x-\imath y,  
\ee 
so that the pair of 2-spinors $\{a,b\}$ represents a static (and spherically symmetric ``in norm'') solution (SSS) of the DE. Selecting then one of the components, say $(b_1)$, of the 2-spinor    $b$ as the initial generating function $F$ instead of (\ref{yukava_a}), we can obtain another solution to the DE of the following form:
\be{yukava_ab1}
a=\left(\begin{array}l
~~~~~~~~0 \\ 
\frac{w}{r^3}(1+mr)e^{-mr}
  \end{array}
  \right), 
 ~b=-\frac{\imath}{mr^5}(3+3mr+m^2r^2)e^{-mr}\left(\begin{array}l
~~w^2\\ 
-wz
  \end{array}
  \right).  
 \ee
Analogously, the choice of the other component $b_2$ of the 2-spinor (\ref{yukava_b}) for the initial generating function $F$ in (\ref{yukava_a}) leads to the following DE solution:
 \be{yakava_ab2}
 a=\left(\begin{array}l
~~~~~~~~0 \\ 
\frac{z}{r^3}(1+mr)e^{-mr}
  \end{array}
  \right), 
 ~b=\frac{\imath}{mr^5}e^{-mr}\left(\begin{array}l
~~~~~~wz(3+3mr+m^2r^2)\\ 
z^2(3+3mr+m^2r^2)+r^2(1+mr)
  \end{array}
  \right).  
 \ee
Evidently, this emerging ``multipole'' chain of solutions is infinite. The components of Dirac fields entering it represent, from the viewpoint of {\it  angular dependence}, various combinations of {\it spherical spinors} (see, e.g.,~\cite[ch.2, sect.11]{Akhiezer}).  Their radial dependence, however, seems to be novel. Indeed, in line with the paradigm of a free electron, one considers, as a rule, only stationary DE solutions of the wave-like type. 

Consider further a {\it stationary} solution to the DE induced by the following spherically symmetric KGE solution:
\be{SSSa}
F=\frac{1}{r} e^{-\imath mt}.
\ee
Taking now the first 2-spinor in the same one-component form $a^T=(0,F)$ and using (\ref{second2sp}) to compute the second 2-spinor $b$, we get
\be{SSSb}
b=\frac{i}{mr^3}e^{-\imath mt}
\left(\begin{array}l
  ~~~~~~w \\
-(z+\imath mr^2)
\end{array}\right).
\ee
From this new solution $\{a,b\}$ of the DE, as from the previous one, one can obtain an infinite chain of the DE-KGE solutions. Remarkably, for solutions of this type the frequency is equal in modulus to the rest mass, $\vert \omega \vert = m$, and the spatial dependence is then determined by the customary {\it Laplace equation}. 

Let us choose now as a generating solution to the KGE a stationary axisymmetrical solution with a ``Dirac string''-type singularity
 \be{string}
 F=\frac{\bar w}{r(r+z)} e^{-\imath mt},~~~\bar w:=x+\imath y.
\ee
Note that, on account of the string-type singularity,  the angular dependence of the solution (\ref{string}) no longer coincides with that typical for spherical spinors. 

Taking again the starting 2-spinor $(a)$ in one-component form  $a^T=(0,F)$ and again using  (\ref{second2sp})  to compute the second 2-spinor, we obtain an axisymmetric ``in norm'' solution (ASS) of the DE~\cite{Equiv}:
\be{stationr}
a=\left(
\begin{array}l
~~~~~0 \\
\frac{\bar w}{r(r+z)} e^{-\imath mt}
\end{array}
\right) , 
~~b=\frac{\imath}{mr^3}e^{-\imath mt} \left(
\begin{array}l
~~~~~z \\
\bar w (1+\frac{\imath mr^2}{r+z})
\end{array} \right). 
\ee
One can now create a chain of DE solutions taking as the starting KGE solution one of the components of the 2-spinor $b$ from (\ref{stationr}). 
The obtained procedure can, evidently, be continued indefinitely. Note, however, that the chain becomes closed if, instead of one solution of the KGE, one will take as a generator the second 2-spinor $(b)$ wholly (see the corresponding example in~\cite{Equiv}). This fact is related to the symmetry of the DE (\ref{Dirac2+2}) with respect to the ``permutation with spatial reflection'' transformation $a(t,\vec r) \leftrightarrow  b(t, -\vec r)$.

To conclude, let us  present two more axisymmetric solutions to the KGE possessing singularities of the string-like type. The first of them has the form   
\be{stereo}
F=\frac{\bar w}{r+z} e^{-\imath mt},  
\ee
so that its spatial part represents the {\it stereographic projection} $S^2 \mapsto C$. This fundamental solution serves for the geometric definition of 2-spinors~\cite{Penrose} and is also especially important in the so-called {\it algebrodynamics}~\cite{AD} where it explicitly corresponds to the Coulomb electric field. The second solution    
\be{suppl_yukava}
F=\frac{\bar w}{r(r+z)} e^{-mr}
\ee
is {\it static} and, in a sense, supplements the SSS (\ref{yukava_a},\ref{yukava_b})  corresponding to the Yukawa potential! These solutions also generate infinite chains of the KGE-DE solutions. In particular, the first of the DE solutions in the chain generated by (\ref{stereo}), has the following form:
\be{stereoDE}
a = \frac{\bar w}{r+z} e^{-\imath m t} \left(
\begin{array}l
0 \\
1
\end{array}
\right), 
~~
b=-\frac{\imath}{mr} e^{-\imath mt} \left( 
\begin{array}l
~~~~~~1 \\
\frac{\bar w}{r+z}(1+\imath mr)
\end{array}
\right), 
\ee 
while the first of the static DE solutions generated by (\ref{suppl_yukava}) looks like
\be{staticDE}
a = \frac{\bar w}{r(r+z)} e^{- m r} \left(
\begin{array}l
0 \\
1
\end{array}
\right), 
~~
b=-\frac{\imath}{mr^3} e^{-mr} \left( 
\begin{array}l
z-mr(r-z) \\
~\bar w (1+\frac{mrz}{r+z})
\end{array}
\right). 
\ee   
Note that the spatial part of (\ref{stereo}), along with the solutions  (\ref{SSSa},\ref{SSSb}) and ASS  ({\ref{stationr}), satisfies the Laplace equation. Making use of other solutions to the latter, one can by differentiation easily obtain new chains of stationary solutions to the DE-KGE.  

\section{Spinors from scalars: non-canonical transformation properties of the Dirac fields}

In the above presentation we considered, as it is generally accepted, both 2-component functions ($a$ and $b$) as 2-spinors. On the other hand, the generating KGE solutions which form, say, the ansatz $a$, according to their internal properties,  should be treated as scalars. Thus, one encounters the problem: how can the scalar nature of the generating  Klein-Gordon fields be matched with the spinor law of transformation of the Dirac field? 

Essentially, this problem has been already solved in~\cite{Equiv}, and below we elucidate the solution on the basis of the 2+2 representation of the DE (\ref{Dirac2+2}) and the above considered examples of field distributions. 

Under (proper) Lorentz transformations of coordinates
\be{lorentz}
X \mapsto \bar X= S X S^+,  
\ee 
where 
\be{coord}
X=X^+ =t+\vec \sigma \vec r
\ee
is the Hermitian matrix of coordinates on $\bf M$, the Weyl operator $W$ (and its conjugate  
 $\tilde W$) are transformed as 
\be{weyltransform}
W \mapsto \tilde S^+ W \tilde S, ~~\tilde W \mapsto  S \tilde W S^+,   
 \ee    
where $S\in SL(2,\mathbb C)$  is a  6-parametric matrix with  ``half-angles'' of (pseudo) rotations as parameters representing (up to a sign) an arbitrary Lorentz transformation from the $SO(3,1)$ group; here $\tilde S, S^+$  are the matrix inverse and the Hermitian conjugate of $S$, respectively. 

According to the canonical spinor law, corresponding to (\ref{lorentz}), transformations of the quantities $a,b$  have the form 
 \be{sptransform} 
 a(X) \mapsto  \bar a (\bar X) =S a(X), ~~~b(X) \mapsto \bar b (\bar X) =\tilde S^+ b( X),
 \ee
so that the DE system (\ref{Dirac2+2}) stays form-invariant, and one obtains a solution to the DE $\{\bar a,\bar b\}$, which is the transform of the original solution $\{a,b\}$ to the new reference frame. 
 
On the other hand, treating the initial components of $a$ subject to the KGE as scalars, one should transform only their arguments, 
 \be{transformarg}
 a (X) \mapsto \bar a (\bar X) =a(X), 
\ee    
after which for the components of $b$, according to (\ref{second2sp}) and (\ref{weyltransform}), one obtains the expression
 \be{transform_nl}
 b(X) \mapsto \bar b (\bar X)= \frac{\imath}{m} \tilde S^+ W \tilde S a(X),
 \ee
so that the pair (\ref{transformarg},\ref{transform_nl}) represents a new solution of the DE $\{\bar a,\bar b\}$ {\it generally distinct} from that canonically transformed (\ref{sptransform}). Note that the new components of $\bar b$, according to  (\ref{transform_nl}), cannot be expressed {\it algebraically} through the initial functions $b$ so that the considered transformations define a {\it nonlinear} representation of the Lorentz group.

In fact,  the possibility of two different types of symmetry transformations of the DE solutions is related to the existence of a supplementary {\it internal} symmetry of the KGE for the doublet of scalar fields $a(X)$ with respect to transformations from the group $SL(2,\mathbb C)_{(INT)}$. The latter is just an independent {\it copy} of the spinor Lorentz group $SL(2,\mathbb C)$ whose transformations do not change the coordinates themselves but linearly intermix the components of the scalar doublet,   
\be{transf_int}
a(X) \mapsto \bar a = M a(X), ~~~M\in SL(2,\mathbb C)_{(INT)}. 
\ee
Combining now these transformations with those from the spinor Lorentz group, we obtain the most general law of transformations for the DE solutions in the following form:
\be{transform_gen}
\bar a (\bar X) = M a (X), ~~~\bar b (\bar X)= \frac{\imath}{m} \tilde S^+ W \tilde S M a(X).
\ee

In general, parameters of the matrix $M$ are entirely independent of those of the Lorentz transformations. On the other hand, if one identifies $M\equiv S$, the canonical spinor law of transformations (\ref{sptransform}) will be restored:
 \be{restor_transform}
 \bar a = Sa(X), ~~~\bar b = \frac{\imath}{m} \tilde S^+ W a(X) \equiv \tilde S^+ b(X).
 \ee
Note, however, that generally, if only the ``half-angles'' parameters of  the matrix $S$ of a 3D rotation do not figure in the matrix $M$, the starting DE solution, transformed  continuously according to (\ref{transform_gen}), {\it returns back to its original values after one complete revolution}; in other words, the {\it customary spinor two-valuedness is generally absent}. From a physical viewpoint, all the KGE-DE solutions obtained from an initial one by means of a combination of Lorentz transformations with transformations of the internal group, should very likely be regarded as {\it equivalent}.
\vskip2mm

Let us now illustrate the above presented scheme on the examples of DE solutions from  section 4. In so doing,  we shall not consider the arbitrary ``intermixings'' of the components (\ref{transform_gen}), but restrict our consideration by two limiting cases: comparing the {\it canonical} spinor transformation (\ref{sptransform})  with  the {\it alternative} (scalar with respect to the components of $a(X)$}) transformation (\ref{transform_nl}). 

It is easy to check that under a rotation of angle $\varphi$ round the $z$-axis, in view of the  axial symmetry, both solutions, SSS (\ref{yukava_a},\ref{yukava_b}) and ASS (\ref{stationr}),  transform rather trivially.  Specifically, according to the canonical law of transformation they acquire the common ``spinor'' factor $e^{\pm \imath \varphi/2}$ (for ASS and SSS, respectively).  On the other hand, under the alternative transformation both components of ASS  get multiplied by the ``quadratic'' phase factor $e^{\imath \varphi}$ and, after a rotation through $360^0$, assume their original values. As for the SSS (defined by the spherically symmetric generating function (\ref{yukava_a})), this solution {\it does not change at all under any 3D rotation} so that one deals with an {\it entirely $SO(3)$- invariant, ``scalar-like''  (with respect to the alternative transformations) solution to the DE} for which $\bar a(X)=a(X),~\bar b(X) = b(X)$! Similarly, the other spherically symmetric ``in norm'' stationary solution (\ref{SSSa},\ref{SSSb}) to the DE is also $SO(3)$-invariant.  

Consider now one more example of a nontrivial transformation of a DE solution for which  the primordial symmetry is broken. For the SSS generated by the spherically symmetric function (\ref{yukava_a}) let us take, as such a  transformation, a Lorentz {\it boost} along, say,  the  $z$-axis with the velocity parameter $V=\tanh{\theta}$. In this case the matrix $S$ takes the form 
\be{matlortransf}
S=\left(
\begin{array}l
e^{-\theta/2}~~ 0 \\    
0 ~~~~~e^{\theta/2}  
\end{array}
\right),
\ee 
and for the SSS transformed {\it canonically}  by  (\ref{sptransform}) we obtain (using for simplicity for {\it new} coordinates the same notation as for the initial ones):
\be{lortransf}
\bar a=\frac{e^{\theta/2}}{r_*}e^{-mr_*}\left(\begin{array}l
0 \\ 
1
\end{array}
  \right),  ~~\bar b = \frac{\imath e^{\theta/2}}{mr_*^3}(1+mr_*)e^{-mr_*}\left(\begin{array}l
~~~~~~~w\\ 
-(1+e^{-2\theta}) z_*
  \end{array}
  \right),  
 \ee
where the following familiar quantities are introduced:
\be{denote}
z_*:=z-Vt,~~~r_*:=\sqrt{x^2+y^2+z_*^2 \cosh^2 \theta}. 
\ee
We see that, apart from the common spinor factor  $e^{\theta/2}$ and transformation of arguments, in the second component of the spinor $b$ there arises a supplementary ``deforming'' factor of the form
\be{deform}
(1+e^{-2\theta})\equiv \frac{1+V}{1-V}.
 \ee
On the other hand, under  the {\it alternative} transformation of the same SSS, in the single component of $a$ only its argument $r$ should be changed to $r_*$ as in (\ref{denote}).  Computing then, explicitly through $\bar a$ or making use of the law of transformation (\ref{transform_nl}), two new components of $\bar b$, one obtains the following solution to the DE in the same moving reference frame:
\be{altlortransf}
\bar a=\frac{1}{r_*}e^{-mr_*}\left(\begin{array}l
0 \\ 
1
\end{array}
  \right),  ~~\bar b = \frac{\imath}{mr_*^3}(1+mr_*)e^{-mr_*}\left(\begin{array}l
~~~~~~~~w \\ 
-(1+e^{-2\theta}) z_*
  \end{array}
  \right),  
 \ee
In contrast with the scalar type of transformation of functions $a$, the corresponding component of $b$ is again deformed by the factor (\ref{deform}), so that the alternatively transformed solution (\ref{altlortransf})  {\it completely reproduces the canonically transformed one} (\ref{lortransf}), disregarding the absence of the common characteristic ``spinor'' factor. Quite similarly are the canonically and alternatively transformed ASS (\ref{stationr})  related to each other under a boost along the $z$-axis. 

Let us note in conclusion that generally, apart from the above considered symmetrical cases, application of the two distinct rules of transformation to an initial DE solution, namely  (\ref{sptransform}) and (\ref{transform_nl}), results in two {\it essentially} different (that is, different not only by the presence/absence of the common spinor factor) solutions to the DE. This fact becomes quite convincing if we consider, say, a rotation of the ASS round an axis which does not coincide with its symmetry axis ($z$).

\section{Two sets of conservative quantities for the Dirac and Klein-Gordon fields} 

As a consequence of the mutual correspondence of the DE-KGE solutions, any Dirac field possesses, apart from the canonical set of integrals of motion, another set of  conserved quantities  defined by the corresponding pair of solutions to the KGE, and vice versa. This  implies, in particular, the existence of a {\it second}, positive definite ``energy''  density for free Dirac fields, as well as positive definite ``probability'' density for any solution of the KGE.  In a more formal way, the above correspondence can be established in the framework of the Lagrangian approach as follows. 

In the 2+2 representation the Dirac equations (\ref{Dirac2+2}) can be obtained through variation of the Lagrangian
\be{lagrDirac}
L = \imath \{a^+ (Wa) +b^+ (\tilde W b) - (Wa^+) a - (\tilde W b^+) b -2m(a^+ b +b^+ a)\}
\ee
(in view of $W^+ = W,~\tilde W^+ = \tilde W$ for the Weyl operator $W$).

Assuming the fulfillment of DE (\ref{Dirac2+2}) and substituting the derivatives in (\ref{lagrDirac}), one arrives at the known property of the Dirac Lagrangian to vanish on the solutions.  On the other hand, replacing in (\ref{lagrDirac}) the fields $a,b,a^+,b^+$ themselves with the derivatives from (\ref{Dirac2+2}), one obtains the Lagrangian for the {\it two doublets} of fields 
\be{lagrKlein}
L=\frac{2}{m} \{(Wa^+) (\tilde Wb) + (\tilde W b^+) (Wa) -m^2 (a^+ b +b^+ a)\}, 
\ee
whose variation results in the KGE for each component of $\{a,b\}$,  
\be{KGE}
(\Box - m^2) a = 0, ~~~(\Box - m^2) b = 0 
\ee
and of their hermitian conjugates. Now making use of the standard procedure, for the original Dirac Lagrangian one defines a canonical set of combinations of field quantities that satisfy on the solutions the {\it continuity equations}. In particular, for the known {\it Dirac current} 4-vector
\be{4-vectorDirac}
j_\mu^{(D)}:= \bar \psi \gamma_\mu \psi,~~~\prt^\mu j_\mu^{(D)} =0, 
\ee
with a positive definite ``charge density'' -- {\it probability density}
 $\rho^{(D)}:=j_0^{(D)} = \psi^+ \psi$, in the 2+2 representation one gets
\be{chargedensD}
\rho^{(D)} =2(a^+a+b^+ b).
\ee      
However, for the same solutions $\{a,b\}$ of the DE-KGE, making use of the Lagrangian (\ref{lagrKlein}), one obtains the expression for a conserved {\it Klein-Gordon current} 4-vector standard for scalar fields:
\be{4-vectorKlein}
j_\mu^{(KG)} =\frac{\imath}{2} (a^+ \prt_\mu a - \prt_\mu a ^+ a + b^+ \prt_\mu b - \prt_\mu b ^+ b), 
\ee    
which defines a {\it sign-indefinite density of the  ``field charge''} 
\be{chargdensKG}
\rho^{(KG)} = \frac{\imath}{2} (a^+ \prt_t a - \prt_t a ^+ a + b^+ \prt_t b - \prt_t b ^+ b).
\ee
One can explicitly observe the difference of these expressions, in particular, for the SSS solution to the DE-KGE (\ref{yukava_a},\ref{yukava_b}) related to the Yukawa potential. Indeed, for this solution the density of the field charge ``{\it a l\' a} Klein-Gordon''   (\ref{chargdensKG}) vanishes while the probability density  (\ref{chargedensD}) is positive and equal to 
\be{staticdens} 
\rho^{(KG)}=\frac{1}{r^2} e^{-2mr}\left(1+\frac{(1+mr)^2}{(mr)^2}\right).
\ee
In  the general case of {\it stationary} solutions to the DE-KGE with $a,b \sim e^{-\imath \omega t}$ the expressions for  the two conserved densities are proportional to each other,  
\be{chargdens_statnr}
\rho^{(D)} =2(a^+a+b^+ b),~~\rho^{(KG)} =2\omega(a^+a+b^+ b),  
\ee
but the sign of the second density can be chosen negative, so that the DE can in fact describe particles with opposite ``charges''. However, for the ASS to the DE (\ref{stationr}) the second density is also positive in view of positivity of the mass $m$.  

Consider now the problem of ``two energies'' for the solutions to the DE-KGE.  Suppose one is given a solution to the DE $\{a,b\}$. Then, making use of the canonical form of the energy-momentum tensor of the Dirac field, for its $(00)$-component, energy density $\epsilon$, in the 2+2 representation one obtains the expression
\be{enrgdensD}
\epsilon^{(D)} = \frac{\imath}{2} (a^+ \prt_t a - \prt_t a ^+ a + b^+ \prt_t b - \prt_t b ^+ b),
\ee
reproducing that for the Klein-Gordon charge density (\ref{chargdensKG})  and, certainly, sign-indefinite.  However, from the corresponding solutions to the KGE and  Lagrangian  (\ref{lagrKlein}) one defines the second ``energy'' density for the same Dirac field: 
\be{enrgdensKG}  
\epsilon^{(KG)}=(\nabla a^+ \nabla a + \prt_t a^+ \prt_t a +m^2 a^+ a) + (\nabla b^+ \nabla b + \prt_t b^+ \prt_t b +m^2 b^+ b), 
\ee
which, of course, is positive definite!
 
Quite analogously, for any DE-KGE solution, apart from the canonical expression, it is possible to define {\it another density of the angular momentum} making use of the expression for the Klein-Gordon field corresponding to the Dirac field. To be sure, this procedure has nothing in common with the generally accepted statement on the one-half spin of Dirac particles or zero spin of particles described by the KGE.
         
\section{On the correspondence of Dirac and Klein-Gordon equations in  electromagnetic field}

One can try to generalize the above obtaned singular DE-KGE solutions for description of fields produced by a point-like or string-like singularity moving along an {\it arbitrary} world line, in a full analogy with the Lienard-Wiehert fields in the massless electromagnetic case. However, already in the case of inertial motion the obtained solutions bring up the de Broglie's interpretation of the wave-particle duality as the concordant motion of a particle-singularity and a ``pilot wave''~\cite{Broglie}. These questions require special consideration.

As for the generalization of the construction  to the external fields, electromagnetic or gravitational, one encounters rather obvious problems here.  The arising obstacles are related to the fact that in these cases the Klein-Gordon operator cannot be factorized to the product of two Dirac operators as in the free case (\ref{factor}). Specifically, in the presence of an electromagnetic (EM) field with 4-potentials  $A^\mu= \{\Phi,\vec A\}$ the {\it squared DE} for the 2-component spinors $a(X), b(X)$ looks as follows (see, e.g., ~\cite[ch.II, sect.12] {Akhiezer} or~\cite{Schweber}):
\be{quadrDirac}
(\Box_{gen}^2 -m^2 +\vec \sigma (\vec H-\imath \vec E) ) a = 0,~~(\Box_{gen}^2 -m^2 + \vec \sigma (\vec H+\imath \vec E) ) b = 0  
\ee 
and, because of the last matrix-valued terms, does not allow for interpretation of the $a$ or $b$ components as scalar fields. In (\ref{quadrDirac}) the first term $\Box_{gen}$ represents the ordinary Klein-Gordon operator in an external field, and $\vec H, \vec E$ -- the fields themselves, magnetic and electric, respectively. 

Remarkably, an attempt to describe the electron-positron field by the squared DE for only one of the 2-spinors has been undertaken in the paper by R.P. Feynman and M. Gell-Mann~\cite{GellMann}. This description is essentially equivalent to the canonical one (see, e.g., ~\cite{Braun}), since the second 2-spinor can be restored through the procedure of differentiation analogous to that above presented. Nonetheless, such a description turns out to be appropriate in the framework of Feynmann formalism of path integration~\cite{Feynman}. 

On the other hand, the scalar nature of the fields generating the DE solutions exhibits itself  when the external (complexified) EM field is {\it (anti) self-dual}. In fact, the (anti) self-duality conditions 
$\frac{\imath}{2}\epsilon_{\mu\nu\rho\lambda} F^{\rho\lambda} =\pm F_{\mu\nu}$ in the 3D-form take just the form
\be{selfdual}
\vec H \pm \imath \vec E =0,~~ (\prt_t \vec A - \nabla \Phi \pm \imath \nabla \times \vec A =0),  
 \ee 
and their fulfillment guarantees the fulfillment of  the homogeneous Maxwell equations, for  the real and imaginary parts of the complex strengths separately ~\footnote{Actually, {\it each} solution of Maxwell equations together with its dual is generated by a complex self-dual solution, see, e.g.,~\cite{IJGMMP,Sing}}. In the case when, say, the conditions of {\it self-duality} $\vec H - \imath \vec E =0$ for potentials (\ref{selfdual}) hold, the squared DE reduces to the ordinary KGE, in an complexified  EM  field, for a doublet of fields $a$ which can then be considered as scalars. These fields after differentiation  (generalizing (\ref{second2sp})) 
\be{gen_b}
 b=\frac{\imath}{m} W_{gen} a,
 \ee
define the second pair of functions $b$ which, together with $a$, form a solution to the DE in the considered field. Here $W_{gen}$  is a generalized, in a usual way (that is, by  ``minimal coupling''), Weyl operator (\ref{weyl}). Note that since the generalized Weyl operators no longer commute, $\tilde W_{gen} W_{gen} \ne W_{gen} \tilde W_{gen}$,  the functions $b$ in (\ref{gen_b}), contrary to $a$,  do not satisfy the KGE in an external field. For {\it antiself-dual} external fields functions $a$ and $b$ exchange places (up to a parity transformation).

The implementation of the above described procedure requires a major reformulation of the canonical problem in relativistic QM concerning the definition of the states of the electron in external EM fields (for a complete review see, e.g.,~\cite{Bagrov}). Particularly, in the relativistic  problem of the hydrogen atom, instead of the standard Coulomb  potential, one should use the combined potential $\{\Phi=q/r,~A_\varphi = \imath \Phi~ \tan(\theta/2), ~~ A_r=A_\theta =0\}$. This results in complex self-dual fields whose real and imaginary parts correspond to electric Coulomb  and magnetic monopole distributions, respectively. Only for such an ansatz does the hydrogen atom problem  reduce to solving the ``pure'' KGE in  the joint  fields of electric and (imaginary) magnetic monopoles~\footnote{In this case the field strengths themselves, contrary to the potentials,  enter only the second equation for $b$ from  (\ref{quadrDirac})}. 

An analogous situation arises in the relativistic problem of the states of particles in a constant and  homogeneous magnetic field. To reduce the DE to the KGE one needs to supplement this field by an {\it imaginary electric} one such that the full complex field be self-dual. Note that this situation can be closely connected with the known problem of imaginary electric dipole moment whose existence inevitably follows from the DE but whose physical meaning still remains unclear. 

A crucial for the present scheme question consists, however, in the distinctions of the {\it energy spectra} of electrons in complex self-dual fields from the canonical ones, in the usually considered real-valued fields. This, fundamental for the considered approach problem requires a thorough investigation. 

As  to the presence of a gravitational field, the situation looks much analogous to the EM case. Specifically, reduction of the DE to the KGE turns out to be possible only in {\it complexified} space-times with complex (anti) self-dual curvature tensor. Such ``(right-) left-flat'' spaces were studied, in particular, in~\cite{Newman} and other papers in connection with the problem of ``nonlinear graviton'', generalizations of twistor theory, etc. Nonetheless, many principal problems arising in this approach,  primarily those about the ways of transition to a real physical metric from the auxiliary complex one, are far from being solved. 
 
\section{Conclusion}

It has been shown above that the DE (at least for free particles) is, essentially, nothing more than an {\it identical interdependency} between derivatives of the doublet of Klein-Gordon fields. In this connection, generating solutions to the KGE manifest themselves as potentials for the Dirac ``field strengths''. One reveals thus that the {\it Dirac field is gauge in nature and closely resembles the Maxwell field}.

We have  established the property of form-invariance of the DE under non-canonical Lorentz transformations in which the generating doublet of Klein-Gordon fields behaves as a pair of {\it scalars} while the second doublet supplementing the first one to a solution of the DE, transforms according to a {\it nonlinear} representation of the Lorentz group. This leads, in particular, to elimination of the generally accepted spinorial 2-valuedness  under  3D rotations. It is known that  the transformation properties of just the {\it free} Dirac/Klein-Gordon fields  predetermine distinct procedures of their secondary quantization in the framework of QED. Therefore, the above presented results  compel one to doubt in the adequacy of the  canonical quantization procedure and ponder over its possible reformulation and/or reinterpretation.   

 In this connection, it is worth noting that A. Zommerfeld ~\cite{Zommerfeld} had proposed another way to regard the Dirac wave functions as scalars instead of generally accepted treating them as bispinors. His proposition makes use of the 4-vector law of transformation of the Dirac $\gamma$-matrices themselves. This procedure preserves  the defining commutation relations (\ref{commut}) for these matrices as well as all the principal consequences of the Dirac theory as a whole (see also ~\cite{Zommer}).    
 
As another consequence of the newly found non-canonical links between the fields of Dirac and Klein-Gordon one can distinguish the possibility to obtain a wide class of (singular) solutions to the DE and KGE by subsequent differentiation of (one or two) starting KGE solutions.  The physical interpretation of thus obtained solutions is generally not evident; however, it would be wrong to ignore their existence. Among these, of special interest is the ``spinor'' analogue (\ref{yukava_a},\ref{yukava_b}) of the mesonic Yukawa potential and the second static solution of the DE (\ref{suppl_yukava}) with a string-like singularity. These solutions can very likely be explicitly interpreted as {\it Dirac fields produced by fermions of two types}. Also remarkable are, certainly, the stationary ``partners''  (\ref{SSSa},\ref{SSSb}), (\ref {stationr}),  (\ref{stereo})  of these static solutions possessing a fixed frequency equal (in appropriate units) to the rest mass. The corresponding chains of solutions to the DE-KGE can be, of course rather speculatively, treated as {\it excited states} (``resonances'').  It seems not be hard to form a complete list of stationary singular solutions to the DE-KGE.   

An interesting variety of these solutions can be obtained by a {\it complex shift}  $z \mapsto z + \imath a$ along the symmetry axis $(z)$ of the original solution. The  thus  deformed solutions (\ref{yukava_a},\ref{yukava_b}) and (\ref{SSSa},\ref{SSSb}) acquire then a {\it ring-like} singularity of radius $a$, as in the case of above-mentioned Appel-Kerr ring~\cite{Witt} corresponding to the Kerr-Newman solution in GTR. As for the solutions (\ref{string},\ref{stationr}), (\ref{stereoDE}) and (\ref{staticDE}), they gain an additional singularity, a straight line string  parallel to the symmetry axis ($z$). It is well known that the Kerr-Newman solution has some properties analogous to those of a Dirac particle (in particular~\cite{Carter}, the gyromagnetic ratio for this solution is $g=2$). Therefore such deformed DE solutions could be very interesting, say, in the context of the {\it Dirac -- Kerr-Newman electron} model developed by A.Ya. Burinskii~\cite{Burin}. 

Finally, it is worth noting that the considered scheme allows to circumvent the Pauli  theorem ~\cite{Pauli} on the sign-indefinite energy density for fields of half-integer spins, and charge density -- of integer spins. Actually, as we have seen, any DE solution can be equipped with a positive definite energy density, and any KGE solution -- with a  positive definite charge density -- probability density~\footnote{Thus, the well-known motivation of Paul Dirac~\cite{DiracMotiv} in his search for a novel relativistic equation with positive definite probability density turns out to be in a sense superfluous!}.

On the whole, however, we do not claim here to suggest some new physical theory, interpretation or quantization procedure. We just intend to simply and rigorously  demonstrate that a number of paradigmatic settings prevailing at present within the (firstly and/or secondary quantized) relativistic field theory are in fact controversial  (questionable) and should actually be reconsidered. For this it is certainly necessary to embark on a number of supplementary investigations, especially for the case of external EM and gravitational fields.

 \end{document}